\documentstyle[aps, preprint]{revtex}
\begin{document}
\preprint{}
\title{Searching for New Physics in the rare decay $B^+ \to D_s^+ \phi $}
\author{Rukmani Mohanta}
\address{School of Physics, University of Hyderabad,
Hyderabad-500 046, India }
\maketitle
\begin{abstract}
The rare decay $B^+ \to D_s^+ \phi $ can occur only via annihilation
type diagram in the standard model. The small branching ratio predicted
in the standard model makes this channel sensitive to new physics
contributions. We analyze this decay mode, both in the
standard model and in
several extensions of it. The models considered are minimal
supersymmetric model with R-parity violation and two Higgs doublet model.
The experimental verification of our findings of large branching ratio
and/or nonzero CP asymmetry may signal the presence of new physics.
\end{abstract}

Recent results from the ongoing experiments in $B$-physics at
BABAR and BELLE have attracted a lot of attention. The main
objective of these $B$ experiments is to explore in detail the
physics of CP violation, to determine many of the flavor
parameters of the standard model (SM) at high precision and to
probe for possible effects of new physics beyond the SM
\cite{ref1,ref2,ref3}. The intensive search for Physics beyond the
standard model is performed now a days in various areas of
particle physics. The $B$ system thus offers a complementary probe
to the search for new physics. In $B$ experiments, new physics
beyond the SM may manifest itself in the following two ways :

{\it i.} decays which are expected to be rare in the SM and are
found to have large branching ratios.

{\it ii.} CP violating asymmetries which are expected to vanish or
to be very small in the SM are found to be significantly large.

Thus the rare $B$ meson decays are suggested to give good
opportunities for discovering new physics beyond the SM. Since their
branching ratios are small in the SM, they are very sensitive to new
physics contributions. In the last few years, different
experimental groups have been accumulating plenty of data for the
rare $B$ decay modes. Some of them have already been measured by
the $B$ factories in KEK and SLAC.

In this context it is interesting to analyze the rare decay $B^+
\to D_s^+ \phi $, which is a pure annihilation type decay
in the SM. The
four valence quarks in the final states $D_s$ and $\phi$ are
different from the ones in the parent $B$ meson, i.e. there is no
spectator quark in this decay. In the usual factorization
approach, this decay mode can be described as the  $\bar b$ and
$u$ quarks in the initial $B$ meson annihilating into vacuum and
the final $D_s$
and $\phi$ mesons are produced from the vacuum afterwards. The
dynamics of exclusive hadronic $B$ decays occurring via the
$W$- exchange or annihilation diagrams, is not yet understood. The
decay rates for such transitions are argued to be negligibly small
due to the suppression of helicity and (or) form factors. However,
a solid justification of this argument is necessary in both theory
and experiments. This decay mode has been recently studied in
perturbative QCD approach with a branching ratio $3 \times
10^{-7}$ \cite{ref4}, which is far below the current experimental
upper limit \cite{ref5}
\begin{equation}
Br(B^+ \to D_s^+ \phi) < 3.2 \times 10^{-4}
\end{equation}
Therefore it provides an appropriate testing ground for physics
beyond the SM.

In this paper we intend to study the decay mode $B^+ \to D_s^+
\phi$, both in the standard model and in several extensions of it.
The models considered here  are the Two Higgs Doublet Model (2HDM)
\cite{ref6} and minimal supersymmetric model with R-parity violation
(RPV) \cite{ref7}.

We first consider the contributions arising from the SM, where
the effective Hamiltonian describing the decay mode is given as
\begin{equation}
{\cal H}_{eff}= \frac{G_F}{\sqrt 2} V_{ub}^* V_{cs}\left [
C_1(\mu)O_1(\mu) +C_2(\mu)O_2(\mu) \right ]
\end{equation}
where \begin{eqnarray}
O_1&=&[\bar b \gamma_\mu(1-\gamma_5) u][\bar
c \gamma^\mu(1-\gamma_5) s]\nonumber\\
O_2&=&[\bar b \gamma_\mu(1-\gamma_5)s][\bar c
\gamma^\mu(1-\gamma_5)u]
\end{eqnarray}
and $C_{1,2}(\mu)$ are Wilson coefficients evaluated at the
renormalization scale $\mu$. In the next to leading logarithmic
approximation their values are evaluated at the $b$ quark mass
scale as \cite{ref8}
\begin{equation}
C_1(m_b)=1.082~~~~~~~~~~{\rm and}~~~~~~~~~~~~C_2(m_b)=-0.185
\end{equation}
The corresponding transition amplitude is given as
\begin{equation}
{\rm Amp}(B^+ \to D_s \phi)=\frac{G_F}{\sqrt 2} V_{ub}^* V_{cs}~a_1
\langle D_s^+ \phi|(\bar c \gamma^\mu(1-\gamma_5) s)(\bar b
\gamma_\mu(1-\gamma_5) u)|B^+ \rangle
\end{equation}
where $a_1=C_1+C_2/N_c$, with $N_C $ is the number of colors. The
evaluation of the matrix element of four fermion operator from the
first principles of QCD is an extremely demanding challenge. To
have some idea of the magnitudes of matrix element, one can still
use the factorization method, factorizing the four quark operators
relevant to non leptonic $B$ decays into the product of two
currents and evaluating separately the matrix elements of the two
currents. Thus in the factorization approximation one can write
the corresponding transition amplitude as
\begin{equation}
{\rm Amp}(B^+ \to D_s^+ \phi)=\frac{G_F}{\sqrt 2} V_{ub}^*
V_{cs}~a_1 \langle D_s^+ \phi|(\bar c \gamma^\mu(1-\gamma_5)
s)|0\rangle \langle 0|(\bar b \gamma_\mu(1-\gamma_5) u)|B^+
\rangle
\end{equation}
The main uncertainties in evaluating the transition amplitude are
due to the matrix element like $\langle D_s^+ \phi|(\bar c
\gamma^\mu(1-\gamma_5) s)|0\rangle $ with $(p_{D_s}+p_\phi)^2
\simeq M_B^2 $. To get an idea about the contributions from
annihilation diagrams, one can assume single pole dominance for
the matrix element and relate it to the crossed channel $\langle
\phi | \bar s \gamma^\mu (1-\gamma_5)c |D_s \rangle $. We see that the
annihilations do not give a large contributions here : with
single pole dominance, form factors are suppressed by a factor
$M_{D_s}^{p2}/(M_B^2-M_{D_s}^{p2})$ because the transferred
momentum $\sqrt{Q^2} =M_B$ is large with respect to pole mass
$M_{D_S}^p$. The annihilation contribution could be larger if
there were other pseudoscalar $(c \bar s)$ resonators
heavier than the pole mass $M_{D_s}^p$, which would enhance the form
factors. The matrix element of the pseudoscalar
and vector meson is usually decomposed
as \cite{ref9}
\begin{eqnarray}
\langle \phi(q, \epsilon)|\bar s \gamma^\mu(1&-&\gamma_5)c| D_s(p)
\rangle = \frac{2 V(Q^2)}{M_{D_s}+M_\phi} \epsilon^{\mu \nu \alpha
\beta}\epsilon_\nu^* p_\alpha q_\beta \nonumber\\
& - &i (\epsilon \cdot Q) \frac{2 M_\phi}{Q^2} Q^\mu \left
(A_3(Q^2) - A_0(Q^2) \right )\nonumber\\
&+&i(M_{D_s}+M_\phi)\left [ \epsilon^{\mu *} A_1(Q^2)
-\frac{\epsilon^* \cdot Q}{(M_{D_s}+M_\phi)^2} (p+q)^\mu A_2(Q^2)
\right ]
\end{eqnarray}
where $Q=(p-q)$ is the momentum transferred during the transition
process. Using the decay constant relation
\begin{equation}
\langle 0 |\bar b \gamma^\mu \gamma_5 u |B^+(Q) \rangle = i f_B
Q^\mu
\end{equation}
the transition amplitude in the SM is given as
\begin{equation}
{\rm Amp}(B \to D_s \phi) = -\frac{G_F}{\sqrt 2}
V_{ub}^* V_{cs}~a_1 f_B~ 2M_\phi A_0(M_B^2)
(\epsilon^* \cdot Q )
\end{equation}
and the corresponding decay width as
\begin{equation}
\Gamma(B \to D_s \phi)= \frac{p_c^3}{8 \pi M_\phi^2}~|
{\rm Amp}(B \to D_s
\phi)/(\epsilon^* \cdot Q)|^2\;,
\end{equation}
where $p_c$ is the c.m. momentum of the decay particles. The value
of the form factor $A_0(0) $ at zero momentum transfer for the
transition $D_s \to \phi $ is given as \cite{ref9} $A_0(0)=0.7$
and the $Q^2$ dependence of the form factor is given as
\begin{equation}
A_0(Q^2)=\frac{1}{1 - \frac{Q^2}{M_{P}^2}} \end{equation} where
$M_P$ is the pole mass with value $M_P=1.97$ GeV \cite{ref9}. Thus we get
\begin{equation}
A_0(M_B^2)=-0.113
\end{equation}
Using $a_1$=1.04 \cite{ch99}, which is extracted from the
experimental data on $B \to D^* (\pi, \rho)$, $f_B$=190 MeV, the
CKM matrix elements from \cite{ref5} and $\tau_B=1.653 \times
10^{-12}s$, we obtain the branching ratio in the SM as
\begin{equation}
Br(B^+ \to D_s^+ \phi)=1.88 \times 10^{-6}
\end{equation}
Thus one can see that the obtained branching ratio in the SM using
the factorization assumption is far below the experimental value
$Br(B^+ \to D_s^+ \phi)<3.2 \times 10^{-4}$. Furthermore, it
should be noted here that the direct CP asymmetry for this decay
mode is {\it zero} in the SM since it receives contribution only
from the single annihilation diagram.

We now proceed to evaluate the branching ratio in the QCD Improved
factorization method, which has been developed recently
\cite{ref10} to study the hadronic $B$ decays. This method
incorporates elements of naive factorization approach as its
leading term and perturbative QCD corrections as subleading
contributions and thus allowing one to compute systematic radiative
corrections to the naive factorization for hadronic $B$ decays.
This method is expected to give a good estimate of the magnitudes
of the hadronic matrix elements in non leptonic $B$ decays.
However, in the QCD factorization approach the weak annihilation
contributions are power suppressed as $\Lambda_{QCD}/m_b$ and
hence do not appear in the factorization formula. Besides power
suppression they also exhibit {\it end point} singularities even
at {\it twist two} order in the light cone expansion of the final
state mesons and therefore can not be computed self consistently
in the context of hard scattering approach. One possible way
to go around the problem is to
treat the end point divergence arising from different sources as
different phenomenological parameters \cite{ref11}. The
corresponding price one has to pay is the introduction of model
dependence and extra numerical uncertainties. In this work we will
follow the treatment of \cite{ref11} and express the weak
annihilation amplitude for $B^+ \to D_s^+ \phi $ as
\begin{equation}
{\rm Amp}(B^+ \to D_s^+ \phi) =\frac{G_F}{\sqrt 2}
V_{ub}^* V_{cs}f_B f_{D_s} f_{\phi}
b_1(D_s, \phi)
\end{equation}
The annihilation parameter $b_1(D_s, \phi)$ is given as
\begin{equation}
b_1(D_s, \phi) = \frac{C_F}{N_c}C_1 A_1^i(D_s, \phi)
\end{equation}
where the color factor $C_F = \frac{N_C^2-1}{2N_C} $ and $N_C=3$.
The function $A_1^i(D_s, \phi)$ is given as
\begin{equation}
A_1^i(D_s, \phi)=\pi \alpha_s\int_0^1dx \int_0^1 dy \psi_{D_s}(x)
\psi_\phi(y) \left [ \frac{1}{y(1-x \bar y)}+\frac{1}{\bar x^2 y}
\right] \;,
\end{equation}
where $\psi_{D_s}(x)$ and $\psi_{\phi}(y)$ are the light cone
distribution amplitudes (LCDA) for the final mesons and $x$ is the
longitudinal momentum fraction of $c$ quark in $D_s$ and $\bar y$
is the momentum fraction of $\bar s$ in $\phi$. Using the
assumption that the LCDAs of the mesons $D_s$ and $\phi $ are
symmetric, one can parameterize the weak annihilation contribution
as \cite{du02}
\begin{equation}
A_1^i(D_s, \phi)\simeq 18 \pi \alpha_s (X_A +\frac{\pi^2}{3}-4)
\end{equation}
where $X_A = \int_0^1 dx/x $ parameterizes the end point
divergence as
\begin{equation}
X_A=\int_0^1 \frac{dx}{x}= {\rm ln} \frac{M_B}{\bar \Lambda}+
\rho e^{-i \theta}
\end{equation}
$\rho $ varies from 0 to 6 and $\theta $  is an arbitrary phase
$0< \theta<360^0 $. Using $\bar \Lambda =\Lambda_{QCD}=200$ MeV
and $\rho e^{i \theta}= i \pi$ as default values along with
$f_B=190$ MeV, $f_{\phi}=233$ MeV, $f_{D_s}=280 $ MeV,
$C_1(m_b)=1.082$ and $\alpha_s(m_b) =0.221$, we obtain the
branching ratio in QCD improved factorization approach as
\begin{equation}
Br(B^+ \to D_s^+ \phi)=0.67 \times 10^{-6}
\end{equation}
We now proceed to calculate the branching ratio for this decay
mode in type II of Two Higgs Doublet Model \cite{ref6}. In type II
2HDM, the up-type quarks get mass from one doublet, while
down-type quarks and charged leptons from the other doublet.
Charged Higgs Yukawa couplings are controlled by the parameter
$\tan \beta = v_2/v_1$, the ratio of vacuum expectation values of
the two doublets, normally expected to be of order $m_t/m_b$. For
our concern the $H^\pm $ effectively induce the four fermion
interaction as
\begin{eqnarray}
{\cal H}_{eff}^{2HDM} &=& -\frac{G_F}{\sqrt 2} \frac{V_{ub}^*
V_{cs}}{m_H^2}
\biggr\{ \bar u \biggr[m_b X (1+\gamma_5) +m_u
Y(1-\gamma_5) \biggr] \biggr\}\nonumber\\
&\times & \biggr\{\bar s \biggr[m_c
Y(1+\gamma_5)+m_sX(1-\gamma_5)\biggr]c \biggr\} \;,
\end{eqnarray}
where $m_H$ denotes the mass of the lightest charged scalar
particle, $m_q$'s denote the constituent quark masses and $X\simeq
1/Y=v_2/v_1=\tan \beta$. In the above equation the terms
proportional to $Y $ can be safely  neglected as $X$ is generally
taken as large \cite{h3}. To evaluate the matrix elements we use
the equation of motion to transform the $(S-P)(S+P)$ currents to
corresponding $(V-A)(V+A)$ form as
\begin{equation}
\bar q_1(1 \pm \gamma_5)q_2= i\left [ \frac{\partial^\mu(\bar q_1
\gamma_\mu q_2)}{m_{q_2}-m_{q_1}} \mp \frac{\partial^\mu(\bar q_1
\gamma_\mu \gamma_5q_2)}{m_{q_2}+m_{q_1}}\right ]\;,
\end{equation}
where the quark masses are current quark masses.
Thus we obtain the expressions for matrix elements as
\begin{equation}
\langle 0 |(\bar b \gamma_5 u)|B^+ \rangle =-\frac{i f_B M_B^2}{m_u+m_b}
\label{eq:w2}
\end{equation}
and
\begin{equation}
\langle \phi |\bar s \gamma_5 c|D_s \rangle =i(\epsilon \cdot Q)
\frac{2 M_\phi}{m_s+m_c} A_0(Q^2)\;.\label{eq:w3}
\end{equation}
Using the above relations, one can obtain the transition amplitude
in 2HDM as
\begin{equation}
{\rm Amp}(B^+ \to D_s \phi )|_{2HDM} = \frac{G_F}{\sqrt 2}
V_{ub}^* V_{cs}\frac{\tan^2 \beta}{M_H^2} f_B\frac{m_b m_s M_B^2}{
(m_b+m_u)(m_s+m_c)}
(\epsilon \cdot Q )2 M_\phi A_0(Q^2)
\end{equation}
So the total amplitude in 2HDM including SM contributions
\begin{equation}
{\cal A}^{SM+2HDM}={\cal A}^{SM}(1-R_1)\;,\label{eq:w1}
\end{equation}
where
\begin{equation}
R_1=\frac{1}{a_1}\frac{\tan^2 \beta}{M_H^2} \frac{m_b m_s M_B^2}{
(m_b+m_u)(m_s+m_c)}
\end{equation}
The free parameters of the 2HDM namely $\tan \beta $, and $M_H$
are not arbitrary, but there are some semi quantitative
restrictions on them using the existing experimental data. The
most direct bound on charged Higgs boson mass comes from the top
quark decays, which yield the bound $M_H > 147 $GeV for large
$\tan \beta $ \cite{h1}. Furthermore, there are no experimental
upper bounds on the mass of the charged Higgs boson, but one
generally expects to have $M_H<1$TeV in order that perturbation
theory remains valid \cite{h2}. For large $\tan \beta $ the most
stringent constraint on $\tan \beta $ and $M_H$ is actually on
their ratio, $\tan \beta /M_H $. The current limits come from the
measured branching ratio for the inclusive decay $B \to X \tau
\bar \nu$, giving $\tan \beta/M_H<0.46 GeV^{-1}$ \cite{h3}. Using
the constituent quark masses as $m_b=4.88 $ GeV, $m_s=0.5 $ GeV
\cite{q1} and current quark masses at the $b$ quark mass scale as
$m_b=4.34$ GeV, $m_c=0.95$ GeV, $m_s=90 $ MeV and $m_u=3.2 $ MeV
\cite{q2}, the branching ratio in 2HDM is found to be
\begin{equation}
Br(B^+ \to D_s^+ \phi)|_{2HDM}=8.0 \times 10^{-6}
\end{equation}

We now analyze the possibility of observing direct CP violation
in this decay mode since it now receives contributions both from the SM
and 2HDM. We can write the decay amplitude (\ref{eq:w1}) as
\begin{equation}
{\cal A}^{SM+2HDM}=|{\cal A}^{SM}|e^{i(\gamma+\delta_1)}
\left [1-|R_1|e^{i(\phi+\delta)} \right ]\;,
\end{equation}
where $\gamma=arg(V_{ub}^*)$ and $\delta_1$ are the weak and strong
phases of standard model amplitudes. $\phi$ and $\delta$ are the
relative weak and strong phases between 2HDM and SM amplitudes.
Thus the direct CP asymmetry for the decay mode is given as
\begin{eqnarray}
a_{cp}&=&\frac{Br(B^+ \to D_s^+ \phi)- Br(B^- \to D_s^- \phi)}
{Br(B^+ \to D_s^+ \phi)+ Br(B^- \to D_s^- \phi)}\nonumber\\
&=&\frac{2 |R_1| \sin \phi \sin \delta}{ 1+|R_1|^2 -2
|R_1| \cos \phi \cos \delta} \label{eq:w4}
\end{eqnarray}
If we set $\phi=\delta= \pi/2$, the maximum possible value of direct
CP violation in 2HDM is found to be
\begin{equation}
a_{cp}(B^+ \to D_s^+ \phi)|_{2HDM} \leq 59 \%
\end{equation}
We now analyze the decay mode in minimal supersymmetric model with
R-parity violation.
 In the supersymmetric models there may be interactions which
violate the baryon number $B$ and the lepton number $L$
generically. The simultaneous presence of both $L$ and $B$ number
violating operators induce rapid proton decay which may contradict
strict experimental bound. In order to keep the proton lifetime
within experimental limit, one needs to impose additional symmetry
beyond the SM gauge symmetry to force the unwanted baryon and lepton
number violating interactions to vanish. In most cases this has
been done by imposing a discrete symmetry called R-parity defined
as $R_p=(-1)^{(3B+L+2S)}$, where $S$ is the intrinsic spin of the
particles. Thus the $R$-parity can be used to distinguish the
particle ($R_p$=+1) from its superpartner ($R_p=-1$). This
symmetry not only forbids rapid proton decay, it also render
stable the lightest supersymmetric particle (LSP). However, this
symmetry is ad hoc in nature. There is no theoretical arguments in
support of this discrete symmetry. Hence it is interesting to see
the phenomenological consequences of the breaking of R-parity in
such a way that either $B$ and $L$ number is violated, both are
not simultaneously violated, thus avoiding rapid proton decays.
Extensive studies has been done to look for the direct as well as
indirect evidence of R-parity violation from different processes
and to put constraints on various R-parity violating couplings.
The most general $R$-parity and Lepton number violating
super-potential is given as
\begin{equation}
W_{\not\!{L}} =\frac{1}{2} \lambda_{ijk} L_i L_j E_k^c
+\lambda_{ijk}^\prime L_i Q_j D_k^c \;,\label{eq:eqn10}
\end{equation}
where, $i, j, k$ are generation indices, $L_i$ and $Q_j$ are
$SU(2)$ doublet lepton and quark superfields and $E_k^c$, $D_k^c$
are lepton and down type quark singlet superfields. Further,
$\lambda_{ijk}$ is antisymmetric under the interchange of the
first two generation indices. Thus the relevant four fermion
interaction induced by the R-parity and lepton number violating
model is
\begin{equation}
{\cal H}_{\not\!{R}} = -
\frac{\lambda_{2i2}^\prime \lambda_{i13}^ {\prime *}} {
4M^2_{\tilde {e}_{L_i}}}~ \bar u (1-\gamma_5)b ~\bar s
(1+\gamma_5) c
\end{equation}
where the summation over $i=1,2,3$ is implied. It should be noted
that the $RPV$ Hamiltonian has the same form as the 2HDM Hamiltonian
except the couplings. So using the Eqs. (\ref{eq:w2}) and (\ref{eq:w3})
one can
easily obtain the amplitude for the $B^+ \to D_s^+ \phi $ in $RPV$
model,
\begin{equation}
{\rm Amp}(B^+ \to D_s^+ \phi)|_{RPV}= \frac{\lambda_{2i2}^\prime
\lambda_{i13}^ {\prime *}} {4M^2_{\tilde {e}_{L_i}}} \frac{f_B
M_B^2}{(m_b+m_u)(m_s+m_c)} (\epsilon \cdot Q)2 M_\phi A_0(Q^2)
\end{equation}
and the total amplitude as
\begin{equation}
{\cal A}^{SM+RPV}={\cal A}^{SM}(1-R_2)\;,
\end{equation}
where
\begin{equation}
R_2=\frac{\sqrt 2}{G_F}\frac{1}{V_{ub}^* V_{cs} a_1}
\frac{\lambda_{2i2}^\prime \lambda_{i13}^{\prime *}}{4M^2_{\tilde
{e}_{L_i}}} \frac{ M_B^2} {(m_b+m_u)(m_s+m_c)}
\end{equation}
Using $\lambda_{2i2}^\prime \lambda_{i13}^ {\prime *}=2.88 \times
10^{-3} $ \cite{ref13}, we obtain the branching ratio in RPV model
as
\begin{equation}
Br(B^+ \to D_s^+ \phi)=3.06 \times 10^{-4}
\end{equation}
Thus we found that the branching ratio in RPV model is quite large
in comparison to SM prediction. So if the experimental value is
found to be in this range, it will definitely be a signal of new
physics beyond the SM.

Replacing $R_2$ in place of $R_1$ in Eq. (\ref{eq:w4}),
the direct CP asymmetry in the decay mode $B^+\to D_s^+ \phi $
in RPV model is found to be
\begin{equation}
a_{cp}(B^+ \to D_s^+ \phi)|_{RPV} \leq 14 \%
\end{equation}

To conclude, in this paper we have calculated the branching ratio
of the two body hadronic decay mode $B^+ \to D_s^+ \phi$ in the
standard model as well as in the two Higgs doublet model and RPV
model. We found that the branching ratio in the RPV model is quite
large in comparison to the SM prediction, whereas the 2HDM
prediction is approximately one order higher than the SM value.
The direct CP asymmetry $a_{cp}$, which is expected to be zero in
the SM, is found to be nonzero and large in both RPV and 2HDM
analyses. From our analyses it follows that the new physics
contribution is quite large and significant. Therefore the rare
decay mode $B^+ \to D_s^+ \phi$ provides an ideal testing ground
to look for new physics.


\begin{thebibliography}{99}
\bibitem{ref1} P. F. Harrison and H. R. Quinn, Editors {\it The Babar
Physics Book}, SLAC-R-504 (1998).
\bibitem{ref2} M. T. Cheng et al, Belle Collaboration, KEK Report 94-2;
F. Takasaki hep-ex/9912004.
\bibitem{ref3} R. Fleischer and J. Matias, Phys. Rev. {\bf D 61}, 074004
(2000).
\bibitem{ref4} C.-D. L\"u, hep-ph/0112127.
\bibitem{ref5} D. E. Groom et al, Review of Particle Physics, Euro.
Phys. J. {\bf C 15}, 1 (2000).
\bibitem{ref6} S. Glashow and S. Weinberg, Phys. Rev. {\bf D 15}, 1958
(1977).
\bibitem{ref7} C. S. Aulakh and R. N. Mohapatra, Phys. Lett.
{\bf B 119}, 136 (1982); F. Zwirner, Phys. Lett. {\bf 132},
103 (1983); I.-H. Lee, Nucl. Phys. {\bf B 246}, 120 (1984);
J. Ellis et al, Phys. Lett. {\bf B 150}, 142 (1985); G. G. Ross
and J. W. F. Valle, Phys. Lett. {\bf 151}, 375 (1985); S. Dawson,
Nucl. Phys. {\bf B 261}, 295 (1985); R. Barbieri, A. Masiero,
Nucl. Phys. {\bf B 267}, 679 (1986): S. Dimopoulos and
L. Hall, Phys. Lett. {\bf B 207}, 210 (1987).
\bibitem{ref8} G. Buchalla, A. J. Buras and M. E. Lautenbacher,
Rev. Mod. Phys. {\bf 68}, 1125 (1996).

\bibitem{ref9} M. Bauer, B. Stech and M. Wirbel, Z. Phys. {\bf 34},
103 (1987).
\bibitem{ch99} H.-Y. Cheng and K.-C. Yang, Phys. Rev. {\bf D 59},
092004 (1999).
\bibitem{ref10} M. Beneke, G. Buchalla, M. Neubert and C. T. Sachrajda,
Phys. Rev. Lett. {\bf 83}, 1914 (1999); Nucl. Phys. {\bf B 591}, 313 (2000).
\bibitem{ref11} M. Beneke, G. Buchalla, M. Neubert and C. T. Sachrajda,
Nucl. Phys. {\bf B 606}, 245 (2000).
\bibitem{du02} D. Du, H. Gong, J. Sun, D. Yang and G. Zhu,
Phys. Rev. {\bf D 65}, 094025 (2002).
\bibitem{h3} ALEPH Collaboration in ICHEP'96, Proceedings of the 28th
International Conference on High Energy Physics, Warsaw, Poland edit.
Z. Ajdeck and A. Wroblewski, World Scientific, Singapore (1997).
\bibitem{h1} F. Abe et. al., CDF Collaboration, Phys. Rev. Lett.
{\bf 79}, 357 (1997).
\bibitem{h2} M. Veltman, Phys. Lett. {\bf B 70}, 253 (1997).
\bibitem{q1} A. Ali, G. Kramer and C.-D. L\"u, Phys. Rev. {\bf D 58}, 094009
(1998).
\bibitem{q2} Y.-H. Chen, H.-Y. Cheng, B. Tseng and K. C. Yang,
Phys. Rev. {\bf D 60}, 094014 (1999).
\bibitem{ref13} D. K. Ghosh, X.-G. He, B. H. J. McKellar
and J.-Q. Shi, hep-ph/0111106.
\end{thebibliography}
\end{document}